\documentclass[ACS,STIX1COL]{WileyNJD-v2}

\usepackage{color}

\usepackage{mathtools}
\usepackage{graphicx}
\usepackage{dcolumn}
\usepackage{array}
\usepackage{lipsum}
\usepackage{bm}
\usepackage{subfigure}
\usepackage{multirow}
\usepackage{tabularx}
\usepackage{amsmath}
\usepackage{braket}
\graphicspath{{plots/}}

\renewcommand{\vec}[1]{\mathbf{#1}}

\newcommand{\beq}{\begin{equation}}
\newcommand{\eeq}{\end{equation}}
\newcommand{\bea}{\begin{eqnarray}}
\newcommand{\eea}{\end{eqnarray}}

\articletype{ORIGINAL ARTICLE}%

\received{xxx}
\revised{xxx}
\accepted{xxx}

\begin{document}

\title{Screening of a test charge in a free-electron gas at warm dense matter and dense non-ideal plasma conditions}

\author[1,2]{Zhandos A. Moldabekov*}


\author[3]{Tobias Dornheim$^\dagger$}


\author[4]{Michael Bonitz$^\mathsection$}

\authormark{Zhandos A. Moldabekov \textsc{et al}}

\address[1]{\orgdiv{Institute for Experimental and Theoretical Physics}, \orgname{Al-Farabi Kazakh National University}, \orgaddress{\state{71 Al-Farabi str.,  
 050040 Almaty}, \country{Kazakhstan}}}
\address[2]{\orgname{Institute of Applied Sciences and IT}, \orgaddress{\state{40-48 Shashkin Str., 050038 Almaty}, \country{Kazakhstan}}}

\address[3]{\orgdiv{Center for Advanced Systems Understanding (CASUS)} \orgname{}, \orgaddress{\state{G\"orlitz}, \country{Germany}}}

\address[4]{\orgdiv{Institut f\"ur Theoretische Physik und Astrophysik}, \orgname{Christian-Albrechts-Universit\"at zu Kiel}, \orgaddress{\state{Leibnizstra{\ss}e 15, 24098 Kiel}, \country{Germany}}}


\corres{*\email{zhandos@physics.kz}\\$^\dagger$\email{t.dornheim@hzdr.de}\\$^\mathsection$\email{bonitz@theo-physik.uni-kiel.de}}


\abstract{
The screening of a test charge by partially degenerate non-ideal free electrons at conditions related to warm dense matter and dense plasmas is investigated using linear response theory and  the local field correction based on \textit{ab inito} Quantum Monte-Carlo  simulations data. The analysis of the obtained results is performed by comparing to the random phase approximation and the Singwi-Tosi-Land-Sj\"olander approximation. The applicability of the long-wavelength approximation  for the description of screening is investigated. 
The impact of electronic exchange-correlations effects on structural properties and the applicability of the screened potential from linear response theory for the simulation of the dynamics of ions are discussed. 
}



\maketitle



\section{Introduction}\label{sec1}

Warm dense matter (WDM) and partially degenerate non-ideal dense plasmas are at the frontier of high-energy-density plasma science \cite{graziani-book, bonitz_pop_19}. These systems are characterized by partial degeneracy of electrons and the importance of inter-particle correlations. 
The last decade witnessed a fast development of the experimental capabilities for creating WDM and dense plasmas in the laboratory, such as laser compression and pulsed power accelerators  \cite{Sinars, hurricane_inertially_2016, Ernstorfer1033},  paving the way to laboratory astrophysics \cite{militzer_massive_2008}. 
On top of it,  the results of the investigation of the fundamental proeprties of high-energy-density plasmas have highly important application in the quest for achieving inertial confinement fusion (ICF) \cite{hu_militzer_PhysRevLett.104.235003, Craxton, Edwards}.  
New experimental data on WDM and partially degenerate dense plasmas have been motivating fast development of theoretical methods and investigations \cite{redmer_glenzer_2009, dornheim_physrep_18, bonitz_pop_20}. Particularly, the progress in Quantum Monte Carlo (QMC) simulations of electrons at WDM and dense plasma parameters ~\cite{dornheim_jcp_19-nn, groth_prb_19, dornheim_prl_18,   groth_jcp17,dornheim_pre17,dornheim_jcp_19,dornheim_pop17,dornheim_prl16,Dornheim_PRL_2020,dornheim_pre_19,Malone_PRL_2016,Malone_JCP_2015,Brown_2014} has allowed a systematic revision and further exploration of various fundamental plasma properties, such as the plasmon dispersion \cite{hamann_cpp_20}, stopping power \cite{moldabekov_pre_20} and thermodynamic properties \cite{dornheim_physrep_18}. Continuing the investigation of the electronic properties in the regime related to WDM and dense non-ideal  plasmas, in this paper we investigate the screening around a test charge (a fixed ion) in a free electron gas computed using the recent neural-net representation of the static local field correction based on \textit{ab initio} QMC simulations \cite{dornheim_jcp_19-nn} within linear response theory.   

There are various simple analytical models (potentials) for the description of a screened potential in plasmas.
The oldest and well known potentials are the Debye-H\"uckel potential, for classical plasmas,  and the Thomas-Fermi potential, for degenerate ideal electrons. 
The advancement of WDM and ICF research has sparked high interest in the study of screening phenomena at partially and weakly degenerate cases \cite{SM_potential, AM_potential, zhandos_cpp17, moldabekov_pop15, moldabekov_cpp15, zhandos_cpp16, zhandos_cpp17, zhandos_cpp_19}. In this regime, the analytic formulas for the  screened potentials were discussed using the long wavelength approximation  and often neglected exchange-correlations effects  \cite{SM_potential, AM_potential, PhysRevE.92.023104,PhysRevE.93.053204,  moldabekov_pop15}.  
Moldabekov \textit{et al} have analyzed various analytical models and the quality of the long wavelength approximation in the WDM regime by comparing them to the RPA result \cite{moldabekov_pop15, moldabekov_pre_18} computed without taking the long wavelength limit as well as to data obtained taking into account electronic exchange correlation effects both in the finite temperature STLS approximation~\cite{tanaka_86} and using ground state QMC data~\cite{Corradini_PRB_1998,Moroni_PRL_1992,Moroni_PRL_1995} for the local field correction \cite{zhandos_cpp17, moldabekov_pre_18}. However, the previous absence of \textit{ab intio} data for the electronic local field correction in the WDM regime had prevented a complete understanding of the impact of  electronic exchange-correlation effects on the screening of an ion  (test charge) in WDM and partially degenerate non-ideal dense plasmas. 

Resently, Dornheim \textit{et al.}~\cite{dornheim_jcp_19-nn}  have developed a machine-learning representation of the static local field correction in the WDM regime based on highly accurate \textit{ab initio} QMC data. In this work, we use this representation to obtain new data for the screened test charge potential. To gauge the role of the exchange-correlations effects,  the results obtained on the basis of the QMC data are compared to the results obtained in the random phase approximation (RPA) where the density response is treated on a mean-field level. Furthermore,  
we compared the QMC-based data to the screened potential computed using an approximate local field correction from the  Singwi-Tosi-Land-Sj\"olander (STLS) approach~\cite{stls_68, 1986JPSJ}.

Additionally, Yukawa-type potentials are often used for the simulations of ionic dynamics in the regime of strong coupling to get an understanding of the basic  physics of phenomena both in WDM and dense plasmas \cite{Ma_PRL,PhysRevE.91.011101, PhysRevLett.116.115003, MURILLO200849, PhysRevResearch.2.033287, PhysRevE.79.010201, zhandos_pre_19}.
Examples of Yukawa pair interaction potential based results for WDM include generalized hydrodynamics~\cite{Mithen}, as well as structural, transport and thermodynamics properties \cite{MURILLO200849, PhysRevResearch.2.033287, Ma_PRL,PhysRevE.91.011101, PhysRevLett.116.115003}. 
Therefore, another important question is to what degree Yukawa type potential  based results are applicable for WDM.
Here, we present the discussion of the  Yukawa type potential obtained using the long wavelength approximation for the electronic density response function with electronic correlations taken into account using  \textit{ab initio} QMC simulations data for the exchange-correlation free energy density\cite{groth_prl17}. In particular, we compare these data to the QMC data based potential obtained  without taking the long wavelength approximation and present results for the structural properties of non-ideal ions obtained using various potentials. 


 The paper is organized as follows: In Sec.~ II, the plasma state of interest and the corresponding dimensionless parameters are defined. 
 In Sec.~III, the theoretical formalism and the methods of calculations are presented. The numerical results are given in Sec.~IV, and in Sec.~V, we summarize our findings.

\section{Dimensionless  parameters}\label{s:2} 

To describe the state of electrons in the WDM regime, we use the   degeneracy parameter $\theta=k_B T_e/E_F$ and the density parameter $r_s=a_e/a_B$, where $a_e=(3/4 \pi n_e)^{-1/3}$ , $E_F$ is the Fermi energy of electrons, $a_B$ is the first Bohr radius, and $n_e$ ($T_e$) is the number density (temperature) of free electrons~\cite{ott_epjd18}. The density parameter is also the non-ideality parameter of the partially or strongly degenerate electrons. WDM and dense non-ideal plasma  states  are characterized by the simultaneous importance of correlation effects (i.e., $r_s\succsim 1$) and partial degeneracy of electrons (i.e., $\theta\sim 1$). Therefore, here we present data for $1\leq r_s \leq 4$ and $0.5\leq \theta \leq 2$. This range of parameters is sufficient to understand the impact of the electronic exchange-correlation effects on screening. On the other hand, the equilibrium  WDM state with a density parameter (of the free electrons) $r_s>4$  and degeneracy parameter $\theta \lesssim 1$  is difficult to realize in experiments, 
due to electron-ion recombinations (see, e.g., the discussion in Refs.~\cite{ bonitz_pop_19,moldabekov_pre_18, Chabrier}). 

Another important parameter for the consideration of screening around a test charge (i.e. an immobile ion) is the coupling parameter between the test charge (ion), $Ze$, and an electron. 
This parameter is introduced as the ratio of the characteristic  Coulomb interaction energy to the  characteristic kinetic energy of electrons for which an estimate is
\begin{align}
    \Gamma(r) = \frac{Ze^2/r}{\left[(k_BT)^2+E_F^2\right]^{1/2} } = \frac{Ze^2}{rk_BT}\frac{1}{\left( 1 + \theta^{-2} \right)^{1/2}}\,,
    \label{eq:gamma}
\end{align}
which, in general, depends on the distance $r$ .

We note that although one usually sets $r=a_e$ in Eq.(\ref{eq:gamma}) \cite{bonitz_pop_20}, we here leave $r$ as a variable. 
The point is that one can find a distance from a point-like test charge (ion) where the coupling parameter becomes greater than one  and vice-versa. 
Thus, we introduce the distance $r_0$ at which $\Gamma(r_0)=1$. At $r>r_0$ we have $\Gamma(r)<1$ and at  $r<r_0$ we have $\Gamma(r)>1$. 
As stated in the introduction, in this paper the screening is computed using linear response theory. Therefore, 
the presented results  are valid at  $r>r_0$ (note that first results for the density response of electrons at WDM conditions beyond linear response theory have recently been reported in Ref.~\cite{Dornheim_PRL_2020}).   

From the condition $\Gamma(r_0)<1$, we find that the linear response theory-based results are reliable  at
\begin{equation}\label{eq:r0}
    \frac{r}{a_B}> \frac{r_0}{a_B} \simeq\frac{Z}{1.84}\frac{r_s^2}{(1+\theta^{2})^{1/2}}.
\end{equation}

In this work, we also discuss the application of the screened potential for the investigation of the physical properties of WDM and dense non-ideal plasmas. 
Therefore, we need dimensionless parameters describing the ionic component of the system. 
In  WDM and dense non-ideal plasma states, ions  usually are strongly correlated and non-degenerate. Therefore, the state of the ionic component is described by the classical coupling parameter:
\begin{equation}
    \Gamma_{ii}=\frac{Z^2e^2}{a_{\rm WS}k_BT_i}, 
\end{equation}
where $a_{\rm WS}=(4/3 \pi n_i)^{-1/3}$ is the Wigner-Seitz radius and $n_i$ ($T_i$) is the number density (temperature) of the ions. 
For simplicity, we consider a two-component system consisting only of one species of ions and free electrons. 

Since experimentally generated WDM is often non-isothermal, i.e., $T_e\neq T_i$, we consider $r_s$, $\theta$, and $\Gamma_{\rm ii}$ as three independent parameters. 
The electron to ion temperature ratio is expressed using these parameters as:
\begin{equation}\label{eq:Te}
    \frac{T_e}{T_i}\simeq 1.84 \frac{\Gamma_{ii}}{Z^{5/3}} \frac{\theta}{r_s}. 
\end{equation}


\section{Theory of screening in the linear response regime}\label{s:parameters}

We consider the screening due to the polarization of the free electrons around a point-like test charge.
The bare potential of the test charge is given by the Coulomb potential, which is considered to be an external field perturbing a uniform electron gas. The perturbation in the electron density leads to an induced potential which compensates the test charge field at sufficiently large distances.   

In the framework of linear response theory, the screened potential of the test charge is expressed using the static dielectric function $\epsilon(\vec k, \omega=0)=\epsilon(\vec k)$ as \cite{Hansen, moldabekov_pre_18}:

 \begin{equation} \label{eq:pot_stat}
 \Phi(\vec r)   = \int\!\frac{\mathrm{d}^3k}{(2 \pi)^3 }~ \frac{4\pi Ze}{k^2}\frac{e^{i \vec k \cdot \vec r}}{\epsilon(\vec k)}   =\frac{Ze}{r}+ \Phi_{\rm ind}(r)\quad,
\end{equation}
where $\Phi_{\rm ind}$ is the induced potential,
\begin{equation}\label{eq:pot_ind}
    \Phi_{\rm ind}(r)=\int\! \frac{\mathrm{d}^3\vec k}{(2 \pi)^3 }  \frac{4\pi Z e}{k^2} \left(\epsilon^{-1}(\mathbf{k})-1\right)  \,\, e^{i \vec k \cdot \vec r} \quad.
\end{equation}

The static dielectric function is expressed in terms of the polarization function,
\begin{align}
    \epsilon(k) &= 1 -  \frac{4\pi e^2}{k^2}\Pi(k),
    \label{eq:df}
\end{align}
where  the static polarization function $\Pi(k)=\Pi(k,\omega=0)$ is computed using the local field correction $G$ and the polarization function in RPA \cite{arista-brandt_84} as the following\cite{Ichimaru}:
   \begin{align}\label{eq:Pi_LFC}
\Pi(k)=\frac{\Pi_{\rm RPA}(k, \omega=0)}{1+\frac{4\pi e^2}{k^2}G(k,\omega=0)\Pi_{\rm RPA}(k, \omega=0)}.
\end{align}

To find the screened potential (\ref{eq:pot_stat}), one then needs to evaluate the integral in Eq.~(\ref{eq:pot_ind}). 
In this work, we analyze the results obtained using a novel QMC-based neural-net representation of the static local field correction \cite{dornheim_jcp_19-nn}. 
To this end, we compare the QMC data based results to the potential computed in RPA [i.e., setting $G=0$ in Eq.(\ref{eq:Pi_LFC})] and to the data obtained using the STLS local field correction~\cite{stls_68, 1986JPSJ}. As it is known \cite{giuliani2005quantum},  the latter has its roots in classical many-particle physics, where the classical two particle distribution function is approximated as $f(\vec r,\vec p;\vec r^{\prime}, \vec p^{\prime}; t)= f(\vec r,\vec p; t) f(\vec r^{\prime}, \vec p^{\prime}; t)g(|\vec r-\vec r^{\prime}|)$,  with $g(|\vec r-\vec r^{\prime}|)$ being the equilibrium electron-electron radial distribution function. 
The STLS scheme is based on the ansatz: 
\begin{equation}
G(\mathbf{k},\omega)\approx G^\textnormal{STLS}(\mathbf{k}) = -\frac{1}{n} \int\frac{\textnormal{d}\mathbf{k}^\prime}{(2\pi)^3}
\frac{\mathbf{k}\cdot\mathbf{k}^\prime}{k^{\prime 2}} [S^{\text{STLS}}(\mathbf{k}-\mathbf{k}^\prime)-1]\;,
\label{G_stls}
\end{equation}
where the static structure factor $S^{\textnormal{STLS}}$ is calculated using the fluctuation-dissipation theorem:
\begin{equation}\label{flucDis}
 S^\text{STLS}(\mathbf{k}) = -\frac{1}{\beta n}\sum_{l=-\infty}^{\infty} \frac{k^2}{4\pi e^2}\left(\frac{1}{\epsilon(\mathbf{k},z_l)}-1\right)\ .
\end{equation}
with $z_l=2\pi il/\beta\hbar$ being the Matsubara frequencies.

Note that to compute the static dielectric function, one can equivalently use the density response function instead of the polarization function (see, e.g., Ref.~\cite{hamann_prb_20}). 


\section{Numerical Results}\label{s:results}


\begin{figure}
    \centering
    \includegraphics[width=0.9\textwidth]{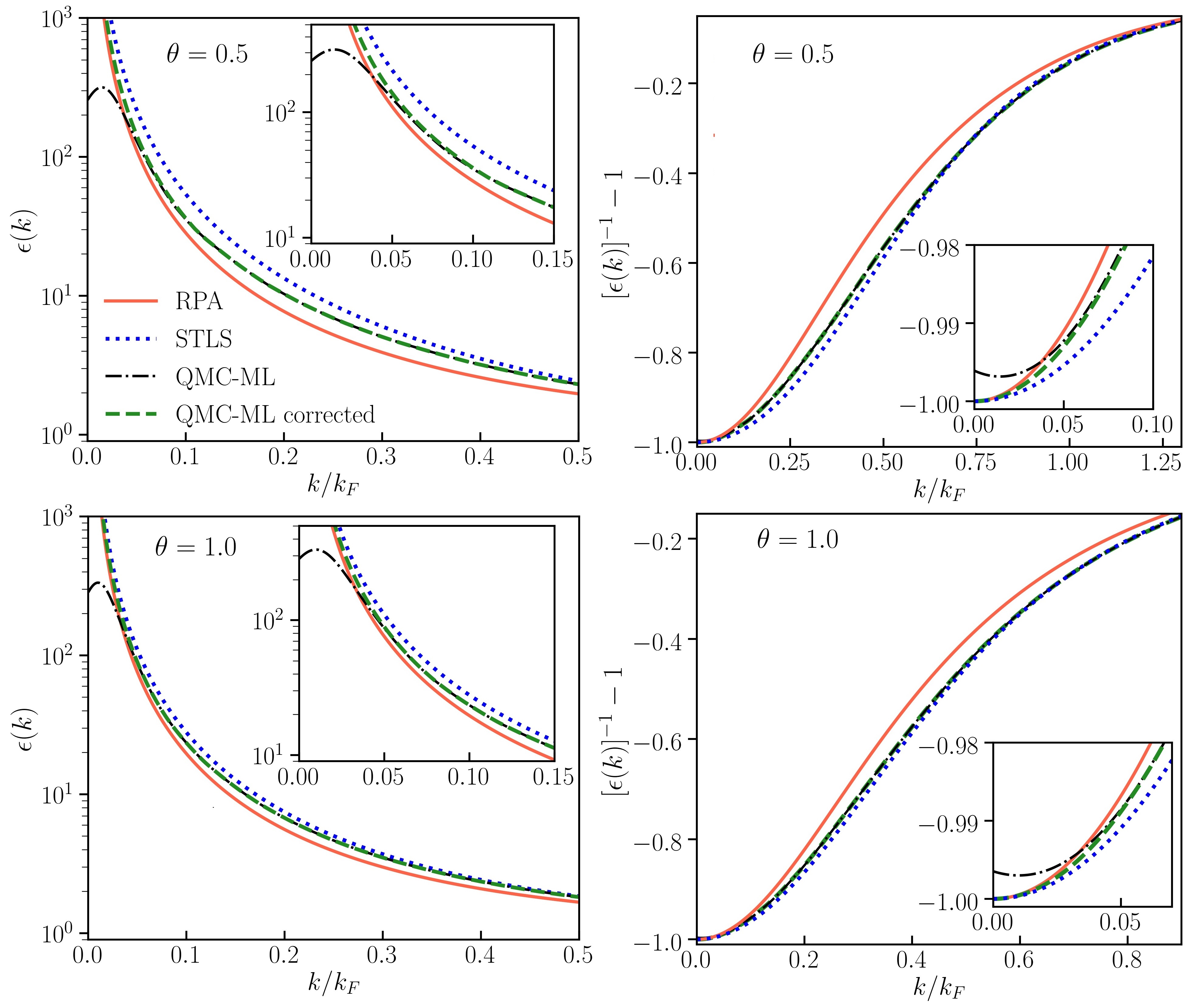}
    \caption{Static dielectric function $\epsilon(k)$ and $[\epsilon(k)]^{-1}-1$ at $r_s=2.0$, and $\theta=0.5$ (top row) and $\theta=1.0$ (bottom row). 
    }
    \label{fig:df}
\end{figure}

\subsection{Static dielectric function}\label{ss:results-epsilon}
To find the QMC data based static dielectric function, we use Eqs.~(\ref{eq:df}) and (\ref{eq:Pi_LFC}), where the local field correction is implemented using the ML representation of the QMC simulations results \cite{dornheim_jcp_19-nn}. The static dielectric function computed in this way is hereafter referred to as QMC-ML results.  
Recently, it has been shown that a straightforward use of the ML representation of the static local field correction leads to incorrect behavior of the static dielectric function at small wavenumbers 
\cite{hamann_prb_20}. 
This is illustrated in Fig.~\ref{fig:df} (left column), where the  QMC-ML data for the dielectric function  is compared to the RPA and STLS results at $r_s=2$, $\theta=0.5$ (top) and $\theta=1.0$ (bottom). We observe that at small wavenumbers  the QMC-ML results tend to a finite value instead of positive infinity with the correct $\sim 1/k^2$ asymptotic. 
The behavior of the QMC-ML data at small wavenumbers was explained by the fact that the ML representation has overall absolute accuracy of $\Delta G\sim0.01$ and that the static local field correction does not  go to zero exactly as $\sim k^2$ in the small wavenumber limit, see Ref.~\cite{hamann_prb_20} for an extensive discussion of this point. 

The problem of the QMC-ML approach can be circumvented using known asymptotic behavior of the static local field correction given by the compressibilty sum-rule~\cite{sjostrom_dufty_2013}:

\begin{equation}\label{eq:G}
 G(k)\simeq \gamma k^2+...,   
\end{equation}
where $\gamma$ is related to the exchange-correlation part of the free energy density $f_{\rm xc}$ as 
\begin{equation}
    \gamma= -\frac{k_F^2}{4\pi e^2}\frac{\partial^2 [n_e f_{\rm xc}(n_e,T_e)]}{\partial n_e^2}.
\end{equation}
In order to enforce the correct behavior of the static dielectric function at small wave numbers, we have extrapolated smoothly the ML representation of $G(k)$  at $k/k_F<0.1$, using  Eq.~(\ref{eq:G}), where $\gamma$ has been computed using the QMC simulations based parameterization of $f_{\rm xc}$ presented in Ref.~\cite{groth_prl17}. The static local field correction computed in this way is then used to find the static dielectric function. The obtained results for $r_s=2$ are included in Fig.~\ref{fig:df} (see the dashed line denoted as  `QMC-ML corrected'), where it is seen that the static dielectric function based on the corrected ML representation of the the local field correction does indeed exhibit the correct behavior at small wavenumbers. 

The computation of the screened ion potential required the numerical evaluation of the integral in Eq.~(\ref{eq:pot_ind}), which functionally depends on $\epsilon^{-1}(k)-1$. Therefore,  we also analyse how the aforementioned behavior of the ML representation of the static local field  correction affects  $\epsilon^{-1}(k)-1$. 
From the right column of Fig.~\ref{fig:df}, we see that  $\epsilon^{-1}(k)-1$ is less sensitive to the spurious behavior of the ML representation of the static local field  correction at small wave numbers. 
At $k\to 0$, the correct limit is $\epsilon^{-1}(k)-1\to -1$. At $r_s=2$,  the QMC-ML results violate the correct limit by less than one percent and start to  deviate from the correct dependence on $k$ at $k\lesssim 0.05 k_F$. In  the right column of Fig.~\ref{fig:df}, the corrected  QMC-ML results for $\epsilon^{-1}(k)-1$ are also shown, which obey the correct asymptotic behavior.
The screened potential computed using the corrected  QMC-ML static dielectric function [denoted `QMC-ML corrected' in  Fig.~\ref{fig:df}] is hereafter referred to simply as \textit{QMC potential}.

\begin{figure}
    \centering
    \includegraphics[width=1.0\textwidth]{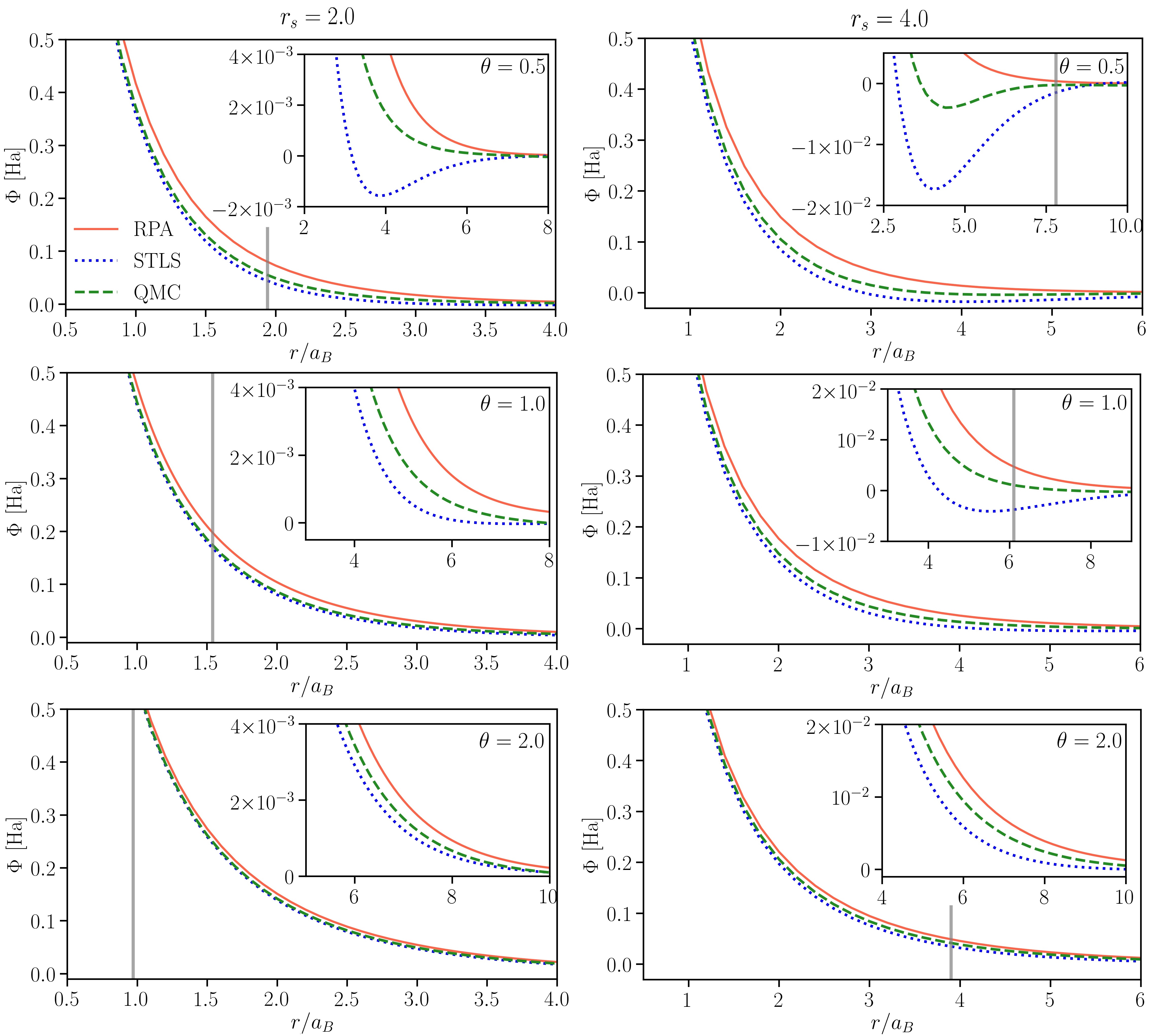}
    \caption{Screened ion potential at $r_s=2$ (left column) and $r_s=4$ (right column) for the different values of the degeneracy parameter $\theta$. The vertical lines indicate the location of $r_0$  computed using Eq. (\ref{eq:r0}). }
    \label{fig:pot-rs2-rs4}
\end{figure}

 Let us conclude this section with a brief discussion of the different results for $\epsilon^{-1}(k)$. For both temperatures, we find that the RPA is systematically too high, by up to $\sim10\%$ as compared to the QMC-based curves and, thus, does not provide a sufficient description of the system even at such a moderate value of $r_s$. In contrast, the STLS exhibits a much better agreement, which is consistent with previous findings for other quantities~\cite{dornheim_physrep_18}.


\subsection{Screened QMC potential}\label{ss:results-potential}

\begin{figure}
    \centering
    \includegraphics[width=1.0\textwidth]{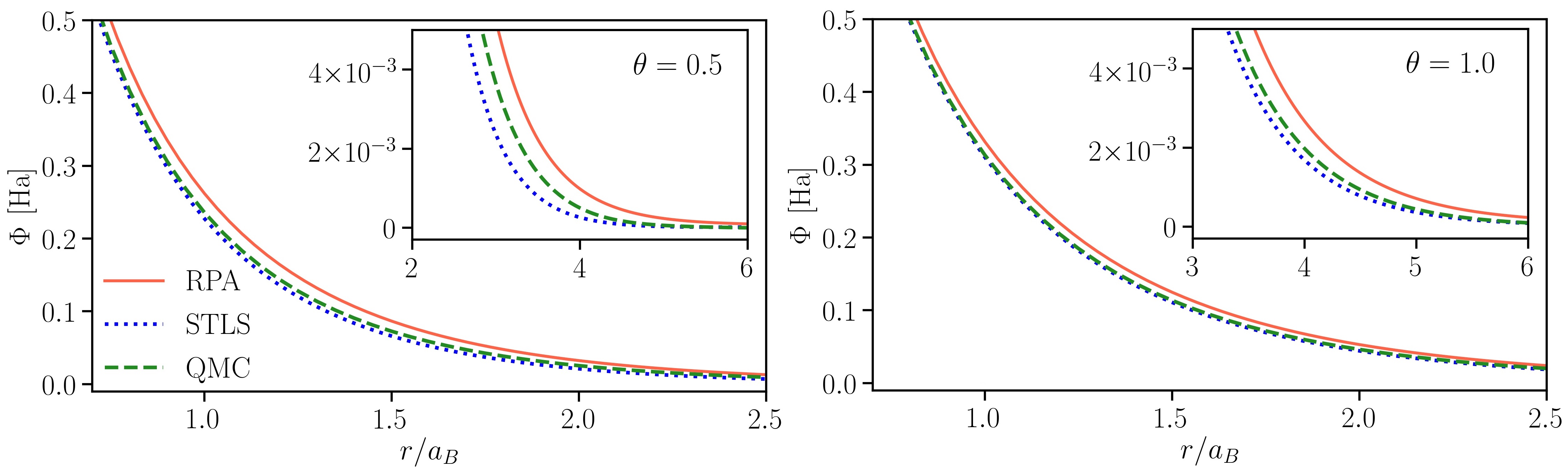}
    \caption{Screened ion potential at $r_s=1$ and two temperatures. Left: $\theta =0.5$, right: $\theta=1.0$. The values  $r_0/a_B\simeq 0.44$, at $\theta=0.5$, and $r_0/a_B\simeq 0.38$, at $\theta=1.0$ are outside of the considered $r$ range.}
    \label{fig:pot-rs1}
\end{figure}

In Figs.~\ref{fig:pot-rs2-rs4} and \ref{fig:pot-rs1} the results for the screened potential computed using Eq.~(\ref{eq:pot_stat}) are presented. In particular,
the QMC based potential is compared to the potential computed neglecting exchange-correlations effects on the density response (\textit{RPA potential}) and with the screened potential obtained using STLS approximation (\textit{STLS potential}). Without  loss  of  generality, the screened potentials are computed for $Z=1$, and the values of the potentials as well as  $r_0$ can be simply rescaled  to  any  other $Z$-value of interest as both of them are linearly proportional to $Z$.

In Fig.~\ref{fig:pot-rs2-rs4}, the data are presented for $r_s=2$ (left column) and $r_s=4$ (right column) at different values of the degeneracy parameter $\theta$. 
The positions of $r_0$ are indicated by the vertical solid grey lines. As discussed in Sec.~\ref{s:parameters}, the presented results are valid to the right of $r_0$.
From  Fig.~\ref{fig:pot-rs2-rs4}, we see that the QMC potential exhibits stronger screening compared to the RPA potential, but weaker screening compared to STLS. For completeness, we mention that such an overestimation of electronic correlation effects within the STLS approximation is consistent with previous findings for different observables, e.g., Refs.~\cite{dornheim_physrep_18,2020arXiv200802165D,Kumar_PRB_2009,giuliani2005quantum}.
With increase in $\theta$, the difference between these potentials diminishes. This is expected, as electronic XC-effects play a less important role for larger temperatures, and, eventually, a mean-field description will be sufficient. The comparison of the results for $r_s=2$ with those for $r_s=4$ shows that the increase in $r_s$ leads to larger difference between the QMC potential and RPA as well as STLS potentials, and, thus, the accurate description of XC-effects becomes even more important.  

From Fig.~\ref{fig:pot-rs2-rs4}, we see that at $r_s=2$, $\theta=0.5$ as well as $r_s=4$, $\theta=0.5$ and $\theta=1.0$, the STLS potential has a negative minimum, which is an artifact of the STLS approximation~\cite{moldabekov_pre_18}. 
This negative minimum, leading to an attraction between like charged ions, fades  with the increase in $\theta$, i.e. the temperature of electrons. 
We note that this behavior of the STLS potential and parameters where it is applicable have been recently discussed by Moldabekov et al.~\cite{moldabekov_pre_18, zhandos_cpp17}. 
At $r_s=4$ and $\theta=0.5$, the QMC potential also develops a very shallow negative minimum at $r\simeq 4.5~a_B<r_0=7.8~a_B$, which is beyond the weak electron-ion coupling distance,  meaning that this result also has to be discarded as a possible artifact of linear response theory. Therefore, even though an exact result for the electronic density response function in the linear response approximation is being used,  the screened potential computed using linear response theory can lead to an unphysical ion-ion attraction at sufficiently large $r_s$ and small $\theta$ (see also Ref.~\cite{Dornheim_PRL_2020} for a first investigation of nonlinear effects of electrons at WDM conditions). 
This question needs further investigation using methods beyond linear response theory, such as Kohn-Sham density-functional, quantum kinetic theory \cite{bonitz-etal.94prb, kwong_prl_00}, or QMC simulation of  an ion embedded in an electron gas or a two-component system.  

As mentioned above, the decrease in the density parameter $r_s$ from 4 to 2 leads to smaller differences between the QMC and STLS potentials. With a further increase of the density to $r_s=1$, the agreement between the QMC and STLS potentials further improves, as can be seen from Fig.~\ref{fig:pot-rs1}. For partially and strongly degenerate states, $r_s=1$ approximately marks the transition from weakly correlated to strongly correlated electrons. 
Therefore, it is interesting to compare the RPA potential to the QMC potential, at this density. From the comparison of the QMC potential to the RPA potential in Fig.~\ref{fig:pot-rs1}, we see that, at $r_s=1$ the electronic exchange-correlation effects are important for the correct description of screening phenomena when $\theta\leq 1$. This conclusion is also supported by the consideration of the screening wavenumber in the next subsection.   

\subsection{Long wavelength approximation}\label{ss:results-dispersion}

\begin{figure}
    \centering
    \includegraphics[width=1.0\textwidth]{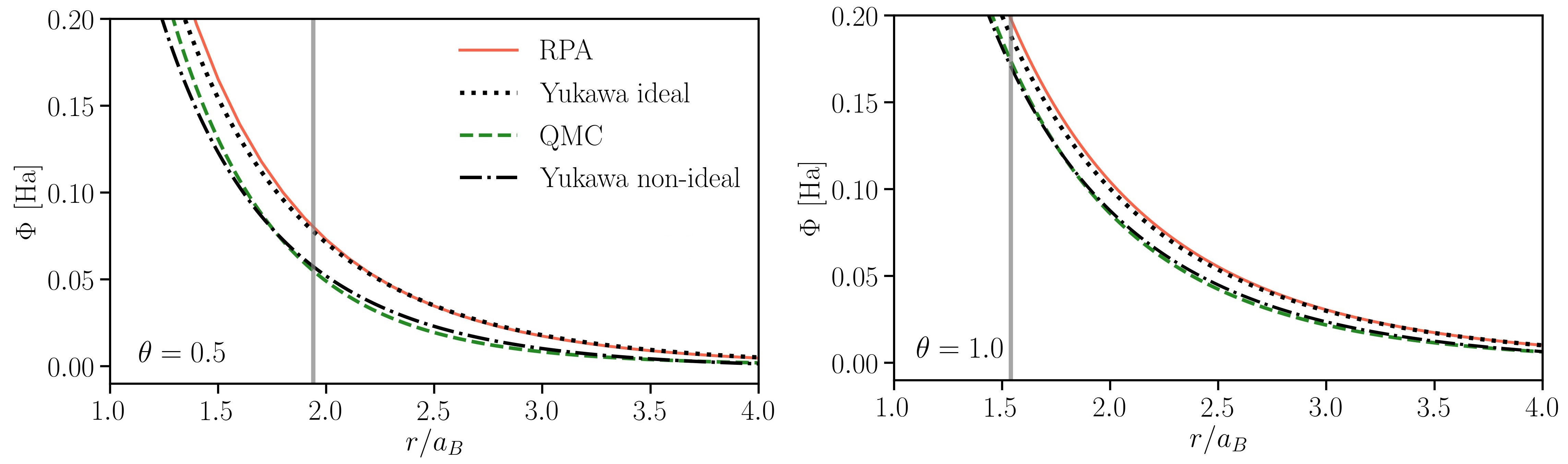}
    \caption{Screened ion potential at $r_s=1$ and two temperatures. Left: $\theta =0.5$, right: $\theta=1.0$. The vertical lines indicate the location of $r_0$  computed using Eq. (\ref{eq:r0}). }
    \label{fig:pot_yukawa}
\end{figure}

Next, we discuss the long-wavelength limit. For that we compare the screened potential computed using  the long-wavelength limit of the dielectric function to the exact QMC potential (i.e., obtained without taking the long-wavelength limit of the dielectric function).

At $k/k_F\ll1$, using the long wavelength expansion of the RPA polarisation function $\Pi_{\rm RPA}^{-1}(k)\simeq 2 a_0$ \cite{moldabekov_pop15, zhandos_pop18} and Eq.~(\ref{eq:G}), we find:
\begin{equation}\label{eq:Pi_app}
    \frac{1}{\Pi(k)}=\frac{1}{\Pi_{\rm RPA}(k)}+\frac{4\pi e^2}{k^2}G(k)\simeq 2a_0+4\pi e^2 \gamma.
\end{equation}
Substituting Eq.~(\ref{eq:Pi_app}) into Eq.~(\ref{eq:df}) and taking into account that $a_0=-2\pi e^2/k_{\rm id}^2$, we recover the well known functional form of the inverse dielectric function in the long wavelength approximation:
\begin{equation}\label{eq:df_2}
    \epsilon^{-1}(k)=\frac{k^2}{k^2+k_s^2},
\end{equation}
\begin{equation}\label{eq:ks}
    k_s^2=k_{\rm id}^2/(1-k_{\rm id}^2 \gamma), \quad k_{\rm id}^2=k_{\rm TF}^2 \theta^{1/2}I_{-1/2}(\eta)/2,
\end{equation}
where $\eta=\mu/k_BT_e$ is the dimensionless chemical potential, $I_{-1/2}$ is the Fermi integral of the order -1/2, and $k_{\rm TF}=\sqrt{3}\omega_p/v_F$ is the Thomas-Fermi wavenumber expressed in terms of the plasma frequency $\omega_p$ and the Fermi velocity $v_F$.    
Using approximation (\ref{eq:df_2}) in Eq.~(\ref{eq:pot_stat}), it is straightforward to obtain the Yukawa-type screened potential with screening length $k_s^{-1}$,
\begin{equation}\label{eq:Yukawa}
    \Phi_Y(r)=\frac{Ze}{r}\exp(-k_s r).
\end{equation}

The RPA case is recovered by setting $\gamma=0$ , i.e. $k_s(\gamma=0)=k_{\rm id}$. This case we call \textit{ideal Yukawa} potential, whereas the general case with $\gamma\neq0$ will be referred to as \textit{non-ideal Yukawa} potential. 
Note that $k_{\rm id}$ interpolates between Debye screening, at $\theta\gg1$, and Thomas–Fermi screening, at $\theta\to0$. 

In Fig.~\ref{fig:pot_yukawa}, at $r_s=2$, $\theta=0.5$ and $\theta=1.0$, we show the comparison of both the QMC and RPA potentials to the respective long-wavelength approximations, i.e, with the non-ideal and ideal Yukawa potentials. As it is expected,  from Fig.~\ref{fig:pot_yukawa}, we see that the presented Yukawa-type screened potentials provide the correct description of the QMC and RPA potentials at large distances. In particular, the agreement with the Yukawa-type screened potential is good at $r>r_0$. This result allows us to generalize our analysis of the impact of the exchange-correlations effect on screening by comparing $k_s$ with $k_{\rm id}$ [as defined by Eq.~(\ref{eq:ks})], i.e., the screening wavenumber computed taking into account electronic non-ideality with that of neglecting exchange-correlations effects. 

\begin{figure}
    \centering
    \includegraphics[width=0.9\textwidth]{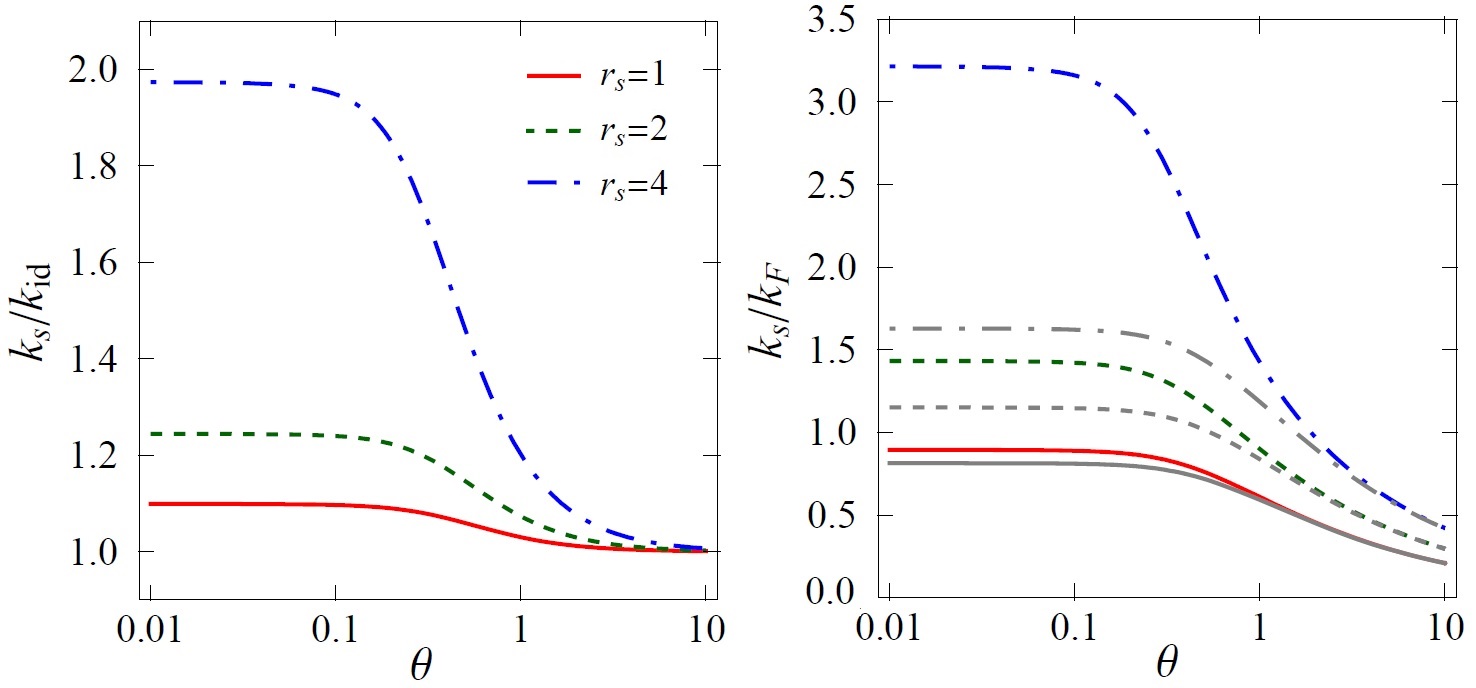}
    \caption{ Screening wavenumbers computed using Eq.~(\ref{eq:ks}) for different values of the density and degeneracy parameters. The left figure presents the ratio  $k_s/k_{\rm id}$. The right figure shows $k_s/k_F$ and $k_{\rm id}/k_F$. }
    \label{fig:ks}
\end{figure}

In the left panel of Fig.~\ref{fig:ks}, we present the results for $k_s/k_{\rm id}$, whereas in the right panel, the comparison of $k_s/k_F$ to $k_{\rm id}/k_F$ is given.
The deviation of the ratio  $k_s/k_{\rm id}$ from unity to larger values indicates stronger screening due to the exchange-correlations effect. 
At $\theta\lesssim 0.1$, the electronic non-ideality leads to about $10\%$ and $25\%$ increase of the screening wavenumber at $r_s=1$ and $r_s=2$, respectively.  At $\theta=1.0$, these numbers decrease to $5\%$ and $10\%$, respectively. At $r_s=4$ and $\theta\lesssim0.1$, the electronic exchange-correlations effect on screening computed within linear response theory results in a significant increase of the screening wavenumber by almost $100\%$. However, from previous discussions we know that at $r_s=4$ and $\theta<1$, linear response theory fails to correctly describe the screened potential. Thus, this significant increase of the screening wavenumber has to be verified by methods correctly capturing the non-linear response of the electrons. At $r_s=4$ and $\theta\simeq 1$, the difference between $k_s$ and $k_{\rm id}$ reduces to about $20\%$. At these parameters, the QMC potential does not exhibit a negative minimum.   A further increase in $\theta$ results in a further weakening of the impact of the electronic non-ideality on screening wavenumber, as it is expected.



\subsection{Effect of electronic correlations on ionic structural properties }\label{ss:ac-results}

\begin{figure}
    \centering
    \includegraphics[width=0.97\textwidth]{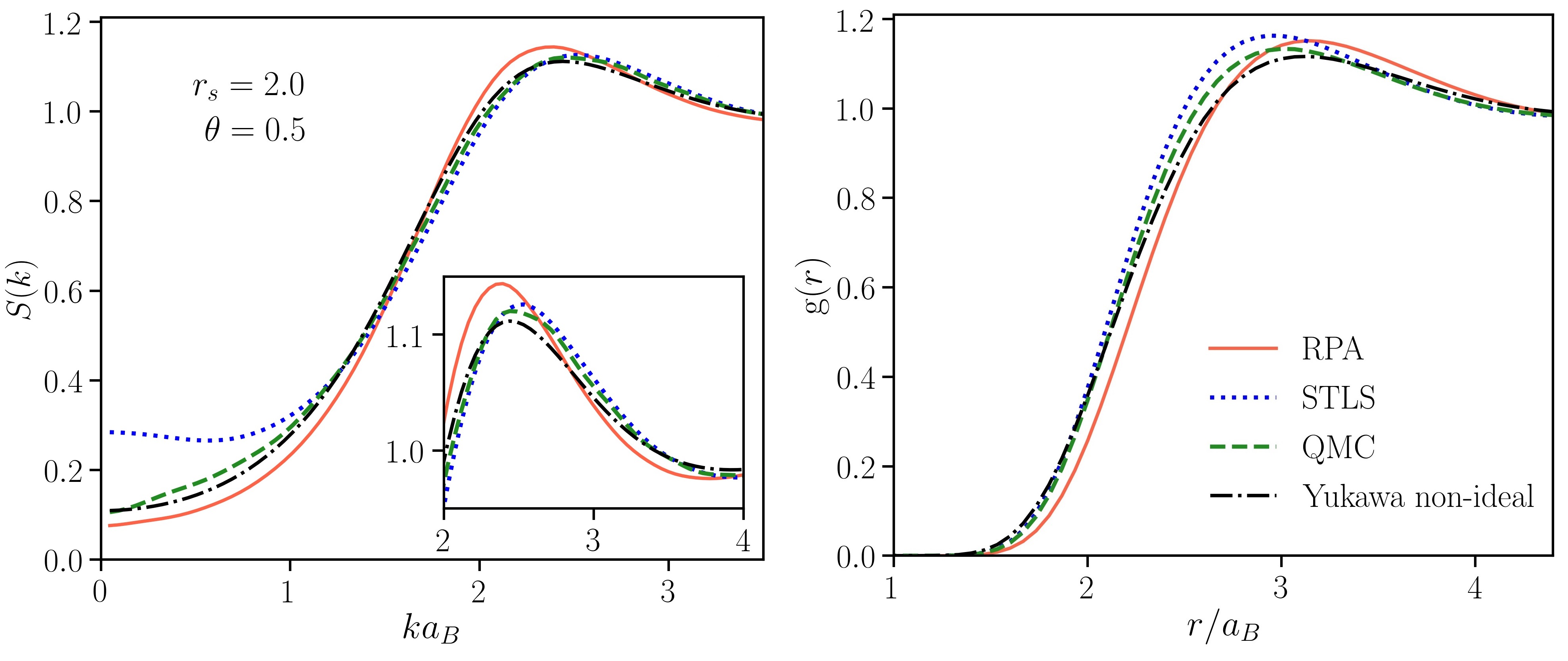}
    \caption{Static structure factor and radial distribution function of strongly coupled ions computed using various screened ion potentials at $\Gamma_{\rm ii}=25$, $r_s=2$ and $\theta=0.5$. }
    \label{fig:SSF}
\end{figure}

To better understand the importance of the discussed difference between the QMC potential and potentials computed using different approximations for the structural properties of the ionic component of a system, the static structure factor (SSF) and radial distribution function (RDF) of ions calculated using different screened ion potentials at $\Gamma_{\rm ii}=25$ are shown  in Fig.~\ref{fig:SSF}.  For illustration, we consider $r_s=2.0$ and $\theta=0.5$. According to Eq.~(\ref{eq:Te}), these parameters correspond to $T_e/T_i\simeq 11$ at $Z=1$ and  $T_e/T_i\simeq 3.5$ at $Z=2$. Such non-isothermal and non-ideal state of ionized matter can be realized experimentally by laser-driven shock-compression \cite{PhysRevLett.98.065002,PhysRevLett.102.115001, moldabekov_pre_18, PhysRevE.94.053211, PhysRevE.92.013103}.
The SSF and RDF presented in Figure \ref{fig:SSF} were obtained by solving the Ornstein-Zernike integral
equation in the hypernetted chain approximation (HNC) \cite{doi:10.1063/1.1679070}. At the plasma parameters considered in Fig.~\ref{fig:SSF}, it was shown in Ref.~\cite{moldabekov_pre_18} that the HNC result is in full agreement with the data computed by molecular dynamics simulation.  

From Fig.~\ref{fig:SSF} we see that, compared to the QMC potential based results, the STLS potential based SSF of ions significantly overestimates SSF values at $ka_B<1$.  The RPA potential based data, in-contrast, underestimate  SSF values at $ka_B<1.5$. Compared to the RPA and STLS potentials based results, the SSF computed using the non-ideal Yukawa potential shows much better agreement with the QMC potential based data at $ka_B<1.5$. At larger wavenumbers $ka_B>1.5$, the non-ideal Yukawa potential based SSF and the STLS potential based SSF exhibit a similar level of accuracy compared to the QMC potential based results. The same is true for the RDF at all distances, while the RPA potential based RDF shows a significant deviation from the QMC potential based RDF for  both the correlation hole at $r/a_B<2.5$ and the peak position of the RDF. In general, the non-ideal Yukawa potential provides better description of the SSF than the RDF. 

The observed impact of the deviation of RPA, STLS and Yukawa type potentials from the QMC potential on the structural properties of the strongly correlated ions becomes stronger with the increase in $\Gamma_{\rm ii}$ and the decrease in $\theta$, as it was discussed recently in Ref.~\cite{moldabekov_pre_18}. Additionally, as mentioned, the increase in $r_s$ leads to a stronger deviation of the STLS and RPA potentials from the QMC potential and as a consequence leads to a stronger disagreement between the corresponding structural characteristics.  

The average distance between ions in the regime of non-ideality ($\Gamma_{\rm ii}>1$) is in the range between $a_{\rm WS}$ and $2~a_{\rm WS}$, as it also can be seen from the RDF curves in Fig.~\ref{fig:SSF}. Therefore, it is clear that  the screened potential obtained using linear response theory can be used for the correct simulation of the dynamics of ions if $r_0<a_{\rm WS}$.
From the condition $r_0<a_{\rm WZ}$, we have:
\begin{equation}
    r_s\lesssim Z^{-4/3} 1.84 (1+\theta^2).
\end{equation}

\begin{figure}
    \centering
    \includegraphics[width=0.6\textwidth]{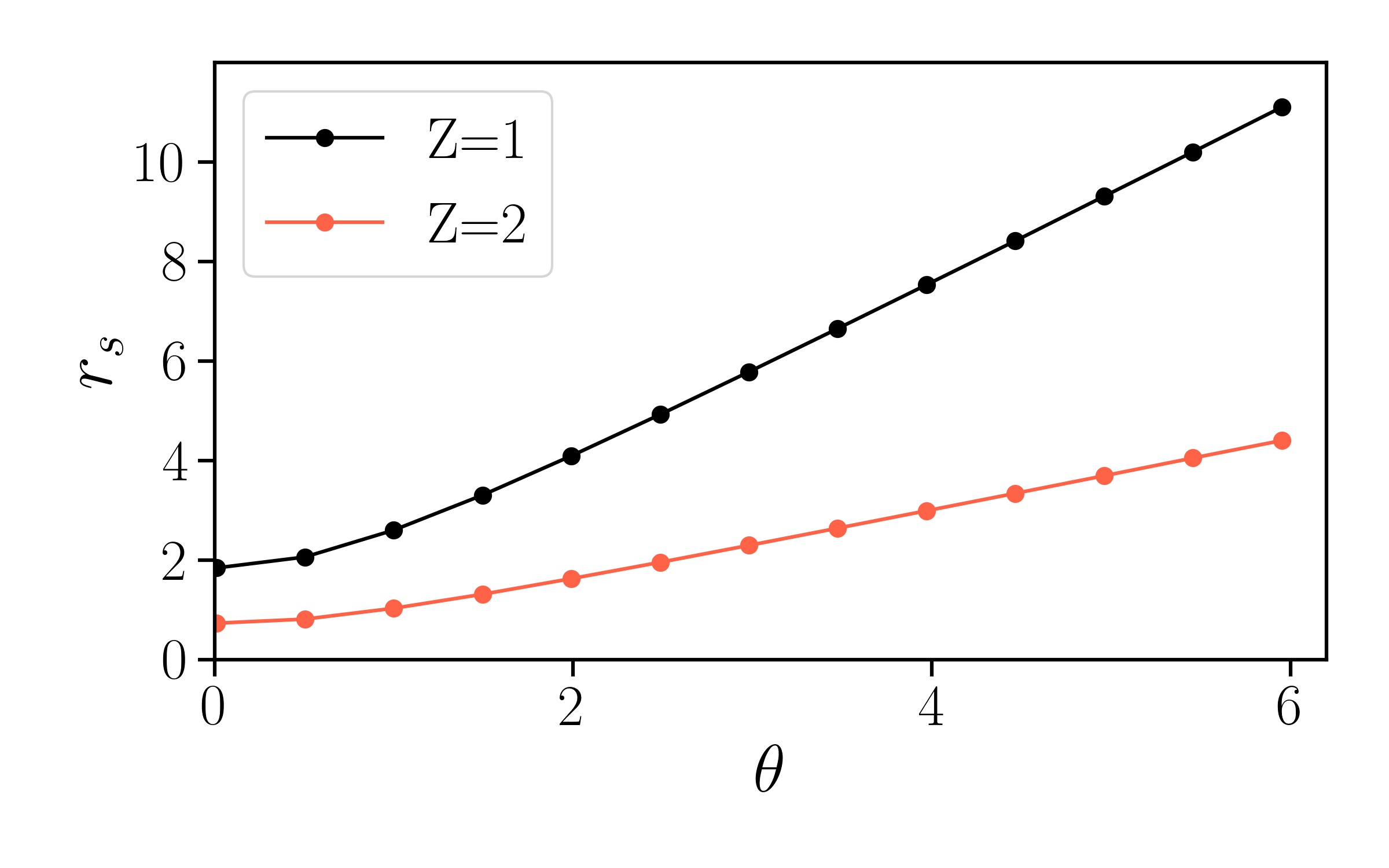}
    \caption{The values of $r_s$ and $\theta$ satisfying the condition $r_0=a_{\rm ii}$ for $Z=1$ and $Z=2$. 
   The linear response result for the screened ion potential can be used for the simulation of a two-component plasma at $r_s$ and $\theta$ values below (to the right of) the
   lines.}
    \label{fig:r0}
\end{figure}

In Fig.~\ref{fig:r0}, the curves of $r_0=a_{\rm WS}$ for $Z=1$ and $Z=2$ are presented. The screened ion potential computed using linear response theory can be used at $\theta$ and $r_s$ values below $r_0=a_{\rm WS}$ curves,   from where, e.g., we see that at $\theta=0.5$ and $Z=1$ the density has to be such that  $r_s\lesssim 2$. 
From Fig.~\ref{fig:r0} we conclude that at the characteristic WDM parameters with $r_s>1$ and $\theta \sim 1$, the applicability of the screened ion potential (based on linear response theory) for the simulation of the ionic component is mainly for the case $Z=1$.

\section{Summary and Discussion}\label{s:summary}

We presented data for the screened ion potential based on linear response theory at WDM and dense non-ideal plasma conditions, where the effect of electronic exchange and correlations was taken into account using a recent neural-net representation of \textit{ab-inito} QMC simulation data. The QMC data-based screened potential has been analyzed by comparing to the results obtained using the RPA and STLS approximations, as well as by considering the long-wavelength approximation, which corresponds to a Yukawa-type potential. 
From this analysis, first of all, we conclude that, compared to the STLS and RPA potentials, the non-ideal Yukawa potential constitutes an overall better approximation to the QMC-based potential and better reproduces the corresponding SSF and RDF of non-ideal ions. 
Secondly,  we have discussed the application of the screened ion potential, computed using linear response theory, for the simulation of the non-ideal ionic component of the system and established that, at $r_s>1$, $\theta\sim 1$ and $\theta<1$, the screened ion potential can be used  only in the case of the singly charged ions.   
Finally, we found that, in the region of the applicability of linear response theory for the description of the ion-ion interaction, electronic exchange-correlation effects lead to an increase (decrease) of the screening wavenumber (wavelength) up to about $25\%$ at $\theta\lesssim 0.1$.

As an outlook, we stress that the non-linear screening regime~\cite{Dornheim_PRL_2020} remains to be explored. Related to this, an open question is the possibility of the ion-ion attraction due to polarization of the electrons around an ion in an equilibrium state in the regime of the strong ion-electron coupling and high ionization degree. Previous predictions of such an effect \cite{shukla_prl_12} were shown to be due to the use of an incorrect dielectric function \cite{bonitz_pre_13,moldabekov_pop15}. The present result is qualitatively different because the involved dielectric function has \textit{ab initio} quality. This strongly hints at fundamental limitations of linear response theory and the crucial importance of nonlinear effects at strong coupling.
Clearly, ion-electron recombination puts certain restriction to the level of the ion-free electron coupling strength. Because of this, as it was established, the characteristic value of the free electron-ion coupling in WDM and dense plasmas is $\Gamma(a)\lesssim 1$ \cite{doi:10.1002/ctpp.201400064}. 

As a final note, we mention that the attraction between like-charged ions is well known in situations out of equilibrium such as in streaming quantum plasmas at $r_s<1$. This manifests itself in wakefield effects~\cite{moldabekov_pre15, ludwig_jpcs_10, zhandos_cpp_19, zhandos_cpp16, moldabekov_cpp15, ludwig_epjd18} that are well known in many fields of plasma physics, including dusty plasmas \cite{block_cpp12,Sundar_2020,PPCF_2020}. The study of wakefields around a projectile in quantum plasmas was, so far, restricted to the case $r_s<1$, due to lack of reliable data for the dynamic local field correction at $r_s>1$. Recently, Dornheim et al.~\cite{2020arXiv200802165D} have developed an effective static approximation for the dynamic local field correction, which provides highly accurate results for electronic properties. This result now allows to extend the study of the dynamical screening in streaming plasmas to the WDM regime with $r_s>1$.

\section*{Acknowledgements}
This work
has been supported by Grant AP08052503 of the Ministry of Education and Science of the Republic of Kazakhstan. 
TD acknowledges support by the Center of Advanced Systems Understanding (CASUS) which is financed by Germany’s Federal Ministry of Education and Research (BMBF) and by the Saxon Ministry for Science, Culture and Tourism (SMWK) with tax funds on the basis of the budget approved by the Saxon State Parliament. 
MB acknowledges support by the German Science Foundation (DFG) via grant BO1366-15.
We further acknowledge CPU time at the Norddeutscher Verbund f\"ur Hoch- und H\"ochstleistungsrechnen (HLRN) under grant shp00015, and on a Bull Cluster at the Center for Information Services and High Performance Computing (ZIH) at Technische Universit\"at Dresden.
%


\bibliography{ref,mb-ref}

\end{document}